\documentstyle[aps,prd,epsfig,theorem]{revtex}
\newcommand{\CO}{{\cal O}}
\newcommand{\lat}{{\rm lat}}
\newcommand{\cont}{{\rm cont}}

\newcommand{\bee}{\begin{equation}}
\newcommand{\ee}{\end{equation}}
\newcommand{\beea}{\begin{eqnarray}}
\newcommand{\eea}{\end{eqnarray}}

\newcommand{\Tr}{\mbox{Tr}}

\theoremstyle{break}    
\theoremstyle{plain}    
\theoremstyle{plain}    
\theoremstyle{plain}    
{\theorembodyfont{\rmfamily}     }
{\theorembodyfont{\rmfamily}     }

\def\g#1{\gamma_{#1}}

\def\sfno#1#2{\overline{(\g{#1}\otimes\xi_{#2})}}

\def\operii{(1\otimes1)}
\def\operix#1{(1\otimes\xi_{#1})}
\def\opergi#1{(\g{#1}\otimes1)}
\def\opergx#1#2{(\g{#1}\otimes\xi_{#2})}

\def\fcskp{\baselineskip 12pt}              
\newcommand{\fc}{\footnotesize \fcskp \advance\itemsep by -6pt}

\preprint{SNUTP-02-025,UW/PT 02-18}
\begin{document}

\title{
  One-loop matching coefficients for improved staggered bilinears}

\author{Weonjong Lee}
\address{
  MS-B285,
  Los Alamos National Laboratory,
  Los Alamos, NM 87544, USA \\
  and \\
  School of Physics,
  Seoul National University,
  Seoul, 151-747, South Korea}
\author{Stephen R. Sharpe}
\address{
  Physics Department, Box 351560
  University of Washington
  Seattle, WA 98195-1560, USA    
  }
\date{\today}
\maketitle
\begin{abstract}
  We calculate one-loop matching factors
  for bilinear operators composed of improved staggered fermions.
  We compare the results for different improvement schemes used in
  the recent literature, all of which involve the use of smeared
  links. These schemes aim to reduce, though not completely eliminate,
  $O(a^2)$ discretization errors.
  We find that all these improvement schemes substantially reduce the size of
  matching factors compared to unimproved staggered fermions.
  The resulting corrections are comparable to, or smaller than,
  those found with Wilson and domain-wall fermions.
  In the best case (``Fat-7'' and mean-field improved HYP links) 
  the corrections
  are $10 \%$ or smaller at $1/a = 2\;$GeV.

\end{abstract}
%
%
%

\section{Introduction}
\label{sec:intro}

Improved staggered fermions are an attractive choice for  
numerical simulations of unquenched QCD~\cite{%
Blum,Lepageproc,DeGrand,Bernard,OrginosToussaint0,Lagae,OrginosToussaint,Lepage,Heller,Orginos,Toussaint}.
They maintain the positive features
of unimproved staggered fermions---smaller CPU requirements
than other fermion discretizations, a remnant chiral symmetry, and
$O(a^2)$ discretization errors---while potentially avoiding its 
drawbacks---large flavor symmetry breaking and large perturbative matching
factors. We have begun a program of calculations of electroweak matrix
elements using these fermions, and thus need to decide between the
different improvement schemes that have been suggested 
in the recent literature. Since we intend at first to use
one-loop perturbation theory to match lattice and continuum operators,
it is important that the matching factors for relevant operators are close to
unity. In this paper we calculate these matching factors for 
the bilinear operators which form the building blocks of the four-fermion
operators we intend to use in our matrix element calculations.
We expect our results to be a good guide to the size of corrections for
the four-fermion operators themselves---typically the one-loop contributions
get roughly doubled. In any case, finding small one-loop corrections for
bilinears is a prerequisite for proceeding to four-fermion operators.

We stress that we are using the term ``improvement'' loosely in this work.
Although the improved actions and operators that we use
are motivated by the Symanzik program,
we are not following this program systematically.
This would involve improving the gauge and fermion actions, and the operators,
so as to remove $O(a^2)$ errors either order by order in perturbation theory
or non-perturbatively.
Such a program is much more difficult for staggered fermions than for 
Wilson-like fermions, particularly with the operators
we use which are spread out over a $2^4$ hypercube.\footnote{%
For recent work on further improving staggered fermions see 
Refs.~\protect\cite{DiPierro,Trottier}.}
Our aim is to change the fermion action
and operators such that the $O(a^2)$ corrections are reduced from the
large size typical of unimproved staggered fermions to the size
seen with Wilson, Domain wall or overlap fermions.
We do not improve the gauge action.

The plan of this paper is as follows. In the next section we
describe the alternatives we have considered for improved operators
and actions. Section \ref{sec:Feynman} collects the new features of the 
Feynman rules that are introduced by improvement. 
In sec.~\ref{sec:analytic}  we present 
analytic results for the one-loop matching constants.
We then, in sec.~\ref{sec:tadhyp},
describe how to do a second level of mean-field improvement of
the bilinear operators.
We close in sec.~\ref{sec:numerical} with the numerical results and
a discussion of their implications.
We collect some definitions in an Appendix, along with
the results that allow us to push the analytic calculation one step further
than in previous work.

We lean heavily on the notation and methodology of Ref.~\cite{PS},
and for brevity we refer to that paper as PS in the following.

\section{Improved actions and operators}
\label{sec:prelim}

Lepage~\cite{Lepageproc,Lepage}
and Lagae and Sinclair~\cite{Lagae}
 have argued that flavor symmetry breaking can be substantially
reduced by suppressing the coupling of high momentum gluons which
connect ``physical'' quarks residing at different corners of the
Brillouin zone.  Such suppression is also expected to
reduce the size of one-loop contributions to perturbative matching
factors, since, as noted by Golterman~\cite{Golterman}, their size is
largely due to tadpole-type diagrams which involve these
flavor-changing vertices.

The flavor-changing coupling can be suppressed by replacing the
standard ``thin'' link with some form of ``smeared'' link in the quark
covariant derivative.  Various options have been tried, and we
consider here two choices which have been successful at reducing
flavor symmetry breaking in pion masses, are relatively local, and
have been extensively studied: the ``Fat-7'' link introduced by
Orginos and Toussaint~\cite{OrginosToussaint}
and studied numerically by the MILC
collaboration~\cite{Orginos}, and the HYP link introduced by
Hasenfratz and Knechtli~\cite{HK}. We refer to the original papers
for the details of the constructions and do not repeat them here.
Both are ``fattened'' by averaging over paths containing links in some
or all of the transverse directions (and which in the Fat-7 case are
up to 7 links long), and in this way they reduce the coupling to
gluons with transverse momenta of $O(1/a)$.  In some sense the Fat-7
link is the simplest choice which accomplishes this, while the HYP
link involves an average over more paths. 
On the other hand, the HYP link involves three levels
of APE-like smearing with projection back into SU(3).
Simulations show that
such smearing is very effective at reducing flavor-symmetry
breaking. Indeed, using the HYP links leads to a greater reduction
in flavor symmetry breaking than the Fat-7 links.

The introduction of smeared links can be viewed as one part of the
Symanzik improvement program applied at tree-level to staggered
fermions.  Complete removal of $O(a^2)$ terms from fermion
vertices requires two other
improvements~\cite{Lepage}.  First, the smearing of the links
introduces an $O(a^2)$ correction to the flavor-conserving quark-gluon
coupling. This can be removed by a adding to the smearing a 5-link
``double staple''---we refer to this as the ``Lepage term''.  Second,
the $O(a^2)$ corrections to the fermion propagator need to be removed,
and this can be done by adding a next-to-nearest-neighbor derivative,
the ``Naik term''~\cite{Naik}.

An observation of practical relevance
is that the Naik term is the only part of the improvement of the
fermion vertex that cannot
be accomplished simply by changing the links in the unimproved staggered
action. In other words, if one does not include the Naik term, and if one
is interested in calculating propagators on configurations that have been
already been generated (whether quenched or unquenched) then the practical
implementation of smeared links is
simple: one calculates the smeared links, and then uses an unimproved staggered
inverter. 

Complete tree-level $O(a^2)$ improvement of physical quantities
requires, in addition to the improvement of fermion vertex outlined above,
the use of a tree-level (or more highly) improved gauge action.
The previous discussion implies, however, that $O(a^2)$ errors from
the gauge action are not responsible for the large flavor-symmetry breaking
or the large ``tadpole'' contributions to matching factors.

With these general comments in mind, we can now explain our choices of
action and operators. 
We use the single-plaquette Wilson gauge action, 
since this is the action we are using in our present simulations.\footnote{%
For this reason, we cannot compare our results with those of
Ref.~\cite{Hein}, since these authors use an improved gauge action.}
For the fermion action, we keep the original staggered
form (without the Naik term), but use various types of smeared links.
The only exception is case (5) below, in which we keep the Naik term.
Finally, for the bilinears we use the standard hypercube form (the
definition of which is given in the Appendix), rendered gauge
invariant by including the average of the product of links along the
shortest paths between the quark and anti-quark field.\footnote{%
We do not consider here so-called ``Landau gauge'' operators---those
rendered gauge invariant by transforming to Landau gauge and then
leaving out the links. These are not useful for matrix elements
involving ``eye'' diagrams, because they allow mixing with
lower-dimension gauge non-invariant operators~\cite{SP}.  They are
also subject to uncertainties due to the presence of Gribov copies.}
The only improvement of the operators that we consider is the use of
smeared links.  Intuitively, the reduction in fluctuations in these
links will reduce the flavor symmetry breaking between
bilinears~\cite{HK}. In all cases we use the same type of smeared
links in the operators as in the action, so that the hypercube vector
current is conserved [except in case (5)].

The specific choices of links we consider are:
\begin{enumerate}
\item
The original gauge links, tadpole improved (following the prescription of
Ref.~\cite{LM} as implemented in PS). We use the fourth root
of the average plaquette 
to determine the ``average link'' $u_0$.
This yields (tadpole-improved) 
unimproved staggered fermions and unimproved operators, and allows
us to check our results against those in PS.
\item
Fat-7 smeared links, built out of tadpole improved links
(as in the numerical implementation of Ref.~\cite{OrginosToussaint,Orginos}).
We stress again that we use these smeared links both in the
action and in the bilinear operators.
\item
Fully $O(a^2)$ improved smeared links, i.e. 
Fat-7 links with the Lepage ``double-staple'' term added,
again both in the action and the operators.
\item
Links smeared according the HYP prescription of Hasenfratz and
Knechtli~\cite{HK}, again both in the action and the operators. 
Three parameters,
$\alpha_{1-3}$, need to be specified to completely define HYP smearing,
and we focus on two choices, as described below.
We also consider a variant in which we tadpole improve the smeared
links themselves (section~\ref{sec:tadhyp}).
\end{enumerate}
In addition, we consider a final choice of action and operators:
\begin{enumerate}
\item [5.]
Following the ``Asqtad'' action introduced by Lepage~\cite{Lepage}
and used extensively
by the MILC collaboration~\cite{Orginos}, we add the
Naik term to the action of choice 3, while taking the same operators
as in choice 3.
In the Naik term alone, we use the original
unsmeared gauge links (tadpole improved). 
\end{enumerate}
This differs from the ``Asqtad'' action, however,
because we use the unimproved Wilson
gauge action, whereas ``Asqtad'' includes an improved
gauge action. We thus refer to our choice as the ``Asqtad-like'' action.
Our expectation is that the choice of gauge action
has relatively little impact on
the size of matching factors, and particularly on the variation
of these factors between bilinears having the same spin and different flavor.

\section{Feynman rules}
\label{sec:Feynman}

The Feynman rules for unimproved staggered fermions are standard.
In the notation we use here, they can be found in PS, 
and we do not repeat them.
We discuss only the changes introduced by smearing the links and
including the Naik term.

We consider first the effect of smearing the links.
For all except tadpole diagrams (i.e. those in which two
gluons emerge from a single, possibly smeared, link), the
only effect is to change the coupling
to the underlying gluon field. With the unimproved action,
a link in the $\mu$-th direction couples only to a gluon $A_\nu(k)$ with
$\nu=\mu$. The smeared links, however, couple to $A_\nu(k)$ for
all $\nu$, and the extra factor this introduces can be conveniently written as
\begin{equation}
\delta_{\nu,\mu} D_\mu(k) + (1 - \delta_{\nu,\mu}) G_{\nu,\mu}(k)
\,.
\label{eq:impvert}
\end{equation}
The diagonal and off-diagonal couplings can be decomposed, respectively, as
\begin{equation}
D_\mu(k) = 1 - d_1 \sum_{\nu\ne\mu} {\bar s}_\nu^2
+ d_2 \sum_{\nu < \rho \atop \nu,\rho\ne\mu}{\bar s}_\nu^2 {\bar s}_\rho^2
- d_3 {\bar s}_\nu^2 {\bar s}_\rho^2 {\bar s}_\sigma^2
- d_4 \sum_{\nu\ne\mu} {\bar s}_\nu^4
\,,
\label{eq:diag}
\end{equation}
with ${\bar s}_\nu= \sin(k_\nu/2)$, etc.,
and
\begin{eqnarray}
G_{\nu,\mu}(k) &=& {\bar s}_\mu {\bar s}_\nu \widetilde G_{\nu,\mu}(k) \\
\widetilde G_{\nu,\mu}(k) &=& d_1 
- d_2 \frac{({\bar s}_\rho^2+ {\bar s}_\sigma^2)}{2}
+ d_3 \frac{{\bar s}_\rho^2 {\bar s}_\sigma^2}{3}
+ d_4 {\bar s}_\nu^2
\,,
\label{eq:offdiag}
\end{eqnarray}
where all indices ($\mu,\nu,\rho,\sigma$) are different.

The coefficients $d_{1-4}$ distinguish the different choices
of links:
\begin{enumerate}
\item
Unimproved (choice 1 above): 
\begin{equation}
d_1 = 0, \quad
d_2 = 0, \quad
d_3 = 0, \quad
d_4 = 0.
\end{equation}
\item
Fat-7 links (choice 2 above):
\begin{equation}
d_1 = 1, \quad
d_2 = 1, \quad
d_3 = 1, \quad
d_4 = 0.
\end{equation}
\item
$O(a^2)$ improved links (choices 3 and 5 above): 
\begin{equation}
d_1 = 0, \quad
d_2 = 1, \quad
d_3 = 1, \quad
d_4 = 1.
\end{equation}
\item
HYP smeared links (choice 4 above):
\begin{equation}
d_1 = (2/3)\alpha_1(1+\alpha_2(1+\alpha_3)), \quad
d_2 = (4/3)\alpha_1\alpha_2(1+2\alpha_3), \quad
d_3 = 8 \alpha_1\alpha_2\alpha_3, \quad
d_4 = 0.
\end{equation}
We consider two choices for the $\alpha_i$. The first was determined
in Ref.~\cite{HK} using a non-perturbative optimization procedure:
$\alpha_1=0.75$, $\alpha_2=0.6$ $\alpha_3=0.3$.
This gives
\begin{equation}
d_1 = 0.89\,, \quad
d_2 = 0.96\,, \quad
d_3 = 1.08\,, \quad
d_4 = 0\,.
\end{equation}
The second is chosen so to remove $O(a^2)$ flavor-symmetry breaking
couplings at tree level. 
This choice, $\alpha_1=7/8$, $\alpha_2=4/7$ and $\alpha_3=1/4$,
gives
\begin{equation}
d_1 = 1, \quad
d_2 = 1, \quad
d_3 = 1, \quad
d_4 = 0,
\end{equation}
i.e. the same as for Fat-7 links.
\end{enumerate}
These results agree with those of Refs.~\cite{Hein,HK2},
but are written here in a somewhat different notation.
The fact that all four choices can be collected in this
form simplifies the resulting one-loop calculations. 
It is particularly noteworthy that the Fat-7 and HYP vertices
can be made identical, showing that these two choices cannot
be distinguished by their flavor breaking effects in 
perturbation theory~\cite{HK2}.
The one-loop matching factors for these two choices are not, however, 
identical, because the tadpole contributions differ.

For tadpole diagrams, which involve two-gluon vertices,
the differences between the actions are more complicated,
and will be given explicitly below. 

The inclusion of the Naik term alters the Feynman rules in several ways.
In the fermion propagator, all factors of $s_\mu=\sin p_\mu$ are
replaced:
\begin{equation}
s_\mu \longrightarrow s^N_\mu = s_\mu (1 + d_N s_\mu^2/6) \,.
\end{equation}
Here we have introduced a fifth coefficient $d_N$ which distinguishes
the different choices of action: $d_N=0$ unless the Naik term is included,
in which case $d_N=1$.
This device allows us to write most of our results in a way which holds for
all choices of action and operators.

The one-gluon vertex is also changed by the Naik term, 
but this can only be represented in a simple way
if one of the quarks in the vertex has vanishing
physical momentum [$k_\mu = (0, \pi/a)$].
In this case, the diagonal part of the vertex 
changes as follows:
\begin{equation}
D_\mu \rightarrow D^N_\mu = D_\mu + d_N s_\mu^2/6 \,.
\end{equation}
This substitution works for all except the self-energy diagram, which
we consider explicitly below.

\section{Analytic results for matching constants}
\label{sec:analytic}

\begin{figure}
\begin{center}
\epsfysize=3in
\epsfbox{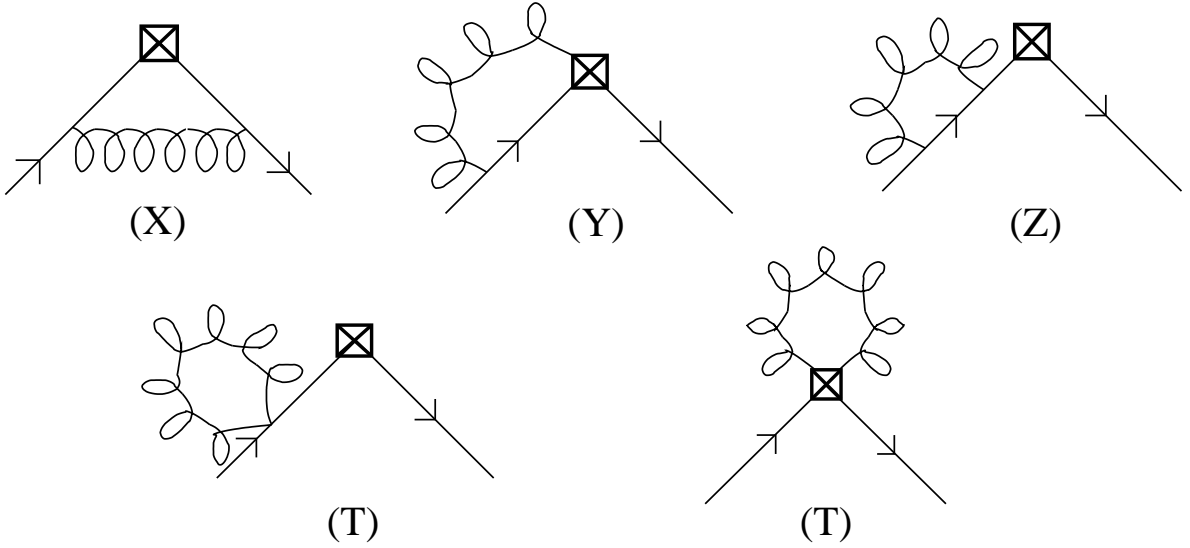}
\end{center}
\caption{Notation for diagrams contributing to matching factors.}
\label{fig:XYZT}
\end{figure}

The one-loop matching relations take the general form,
\begin{equation}
\CO_i^{\cont} = \CO_i^{\lat} +
C_F {g^2\over 16\pi^2} \sum_j
(\delta_{ij} 2 d_i \ln(\mu a) + c_{ij}) \CO_j^{\lat}
\label{eq:matching}
\end{equation}
where $C_F=4/3$ is the color Casimir factor,
$\mu$ is the renormalization scale of the continuum operators,
and $i$ and $j$ run over all the different possible bilinears in a four-flavor
theory. The explicit forms of the operators are given in the Appendix.
The constants $d_i$ are proportional to the one-loop anomalous dimensions
of the bilinears, $\gamma_i^{(0)} = - 2 C_F d_i $.
They depend only on the spin of the bilinear, and are $d_i= (3,0,-1)$
for spins $(S/P,V/A,T)$.
The finite part of the coefficient can be written
\begin{equation}
c_{ij} = \delta_{ij}\left[
d_i (\gamma_E - F_{0000}) + t_S\right] - c^1_{ij}
\,,
\label{eq:cij}
\end{equation}
with $t_S$ depending on the continuum renormalization scheme.
For the NDR scheme $t_S=(-0.5,0,1.5)$ for spins $(S/P,V/A,T)$.
The conversion to other schemes is given in PS.
The constants are $\gamma_E=0.577216 $ and $F_{0000}= 4.36923$.
Finally, the ``lattice'' part of the coefficient can be broken
up as follows:
\begin{equation}
c^1_{ij} = X_{ij} + \delta_{ij}\left(Y_i + T_i + Z\right)
\,.
\end{equation}
Here $X$, $Y$, $T$ and $Z$ refer to contributions from
the different types of diagrams using the notation of PS,
as illustrated in Fig.~\ref{fig:XYZT}. 
This equation incorporates the fact, derived below, that only
the $X$ diagrams lead to mixing among bilinears.

All our calculations are done in Feynman gauge. 
We have checked our results by doing two independent calculations
using different methods---the first following PS and the
second using the methods presented in Refs.~\cite{Leepert,LeeKlomfass}. 

\subsection{$X$ diagrams}
\label{sec:X}

The calculation follows the same steps as in PS, except for
two changes.
\begin{itemize}
\item
We have been able to carry out the calculation
analytically apart from one final integral,
using the results (\ref{eq:vertex},\ref{eq:mixingvertex}) given in 
the Appendix.
\item
The improved vertex eq.~(\ref{eq:impvert})
allows propagation from a smeared link
in the $\mu$-th direction to another in any direction, even in Feynman gauge.
It is useful to distinguish between the case where the second link
is also in the $\mu$-th direction, for which the
gluon propagator is multiplied by
\begin{equation}
P^{NN}_{\mu}(k) = D^N_\mu(k)^2 + \sum_{\nu\ne\mu} G_{\nu,\mu}(k)^2
\,,
\end{equation}
and the case where the second link is in a different direction $\rho\ne\mu$,
for which the multiplying factor is
\begin{equation}
D^N_\mu G_{\mu,\rho} + D^N_\rho G_{\rho,\mu}
+ \sum_{\nu\ne(\mu,\rho)} G_{\nu,\mu}G_{\nu,\rho}
\equiv \ 4 \bar s_\mu \bar s_\rho O^{NN}_{\mu\rho}(k)
\,.
\end{equation}
The superscripts emphasize the fact that there is a possible Naik term at both
ends of the propagator.
\end{itemize}

Using these results, we find the following expression
for the diagonal part of the
contributions from the $X$ diagrams:
\begin{eqnarray}
X_{ii} &=& \sum_{\mu,\rho} \int_k
\left[ \bar c_\mu^2 P^{NN}_{\mu} (s^N_\rho)^2 B F^2 V_i(k)
	- \frac{B^2}{4} \right]
(-1)^{\widetilde S_\mu + \widetilde S_\rho}\nonumber \\
&&\mbox{}
+ 2\sum_{\mu<\nu} \int_k
s_\mu s^N_\mu s_\nu s^N_\nu O^{NN}_{\mu\nu} B F^2 V_i(k)
\left[1 - (-1)^{\widetilde S_\mu + \widetilde S_\nu}\right]
\,. \label{eq:Xii}
\end{eqnarray}
Here 
$\int_k \equiv 16\pi^2 \prod_\mu \int_{-\pi}^{\pi} {dk_\mu \over 2\pi}$,
$\bar c_\mu=\cos(k_\mu/2)$ and $s_\mu=\sin(k_\mu)$.
The functions arising from boson and fermion propagators are,
respectively,
\begin{equation}
B(k) = {1\over 4 \sum_\mu \bar s_\mu^2} \,,\qquad
F(k) = {1\over \sum_\mu (s^N_\mu)^2} \,.
\end{equation}
For the sake of brevity we do not show the argument $k$ of these
functions or of $P^{NN}_\mu$ and $O^{NN}_{\mu\nu}$ in eq.~(\ref{eq:Xii})
and in the following.
The index $i$ in eq.~(\ref{eq:Xii}) labels the spin and flavor of the operator,
and the $B^2/4$ term  on the first line
is the conventional integral used to cancel divergences.
The function $V_i(k)$ is defined in eq.~(\ref{eq:vertex});
we stress again that the use of this equation leads to a simpler
form than that given in PS.

Only the $X$ diagrams lead to mixing, i.e. non-zero values for
$X_{ij}$, $i\ne j$.
We find that we can also give explicit expressions for the mixing terms
using eq.~(\ref{eq:mixingvertex}).
As for unimproved staggered fermions, the mixing that occurs at one-loop
turns out to be only a subset of that allowed by the hypercubic symmetry group.
The non-zero mixing coefficients are 
(using the definitions in PS---see table~\ref{tab:cij})
\begin{eqnarray}
c_{VVM} &=& \int_k 4 B F^2 s_1 s^N_1 s_2 s^N_2 
		 (P^{NN}_3 \bar c_3^2 - O^{NN}_{12} s_3^2) 
\,,\\
c_{VAM} &=& \int_k 4 B F^2 s_1 s^N_1 s_2 s^N_2 
	(-P^{NN}_3 \bar c_3^2 c_3 c_4 + O^{NN}_{12} s_3^2[2c_2 - c_3]c_4) 
\,,\\
c_{VTM} &=& \int_k 2 B F^2 s_1 s^N_1 s_2 s^N_2 
	(-P^{NN}_3 \bar c_3^2 [c_3+c_4] + O^{NN}_{12} s_3^2[2c_2 - c_3 +c_4]) 
\,,\\
c_{TAM} &=& \int_k 2 B F^2 s_1 s^N_1 s_2 s^N_2 
	(P^{NN}_3 \bar c_3^2[c_4-c_3] + O^{NN}_{12} s_3^2[2c_2 -c_3-c_4])
\,,
\end{eqnarray}
where $c_\mu=\cos(k_\mu)$.

\subsection{$Y$ diagrams}
\label{subsec:Y}

$Y$ diagrams involve the gluon connecting an external quark or antiquark line
to the operator.
As explained in PS,
with the unimproved staggered action and unsmeared links,
$Y$ diagrams do not lead to mixing between different bilinears,
and the result depends only on the ``distance'', 
$\Delta=\sum_\mu (S-F)_\mu^2$,  of the 
bilinear. It is straightforward, though tedious, to check that these arguments
generalize to the improved actions considered here. The result is
\begin{equation}
Y_\Delta = Y_{\Delta-1} + I_\Delta \qquad (\Delta > 0)
\,,
\end{equation}
with $Y_{\Delta=0}=0$, and
\begin{eqnarray}
I_\Delta &=& \int_k B F s_1 s^N_1 P^N_{1}V_Y(\Delta)
+ \int_k 12 B F\bar s_1^2 s_2 s^N_2 O^N_{21} V_Y(\Delta) \\
V_Y(1) &=& 1 \,,\ \ 
V_Y(2) = \frac{c_2+c_3+c_4}{3} \,,\ \ 
V_Y(3) = \frac{c_2 c_3 + c_2 c_4 + c_3 c_4}{3} \,,\ \ 
V_Y(4) = c_2 c_3 c_4 \,.
\end{eqnarray}
The new functions are defined by
\begin{equation}
P^{N}_{\mu}(k) \equiv D^N_\mu(k) D_\mu(k) 
+ \sum_{\nu\ne\mu} G_{\nu,\mu}(k)^2
\,,
\end{equation}
and
\begin{equation}
4 \bar s_\mu \bar s_\rho O^{N}_{\mu\rho}(k) \ \equiv\ 
D^N_\mu G_{\mu,\rho} + D_\rho G_{\rho,\mu}
+ \sum_{\nu\ne(\mu,\rho)} G_{\nu,\mu}G_{\nu,\rho}
\,.
\end{equation}
The single superscript ``$N$'' reflects the fact that the Naik term appears
only at the quark-gluon vertex and not in the operator.
Note that, unlike $O^{NN}_{\mu\rho}$, $O^N_{\mu\rho}$ is not symmetrical.

\subsection{Tadpole diagrams}
\label{subsec:tad}

Here we include tadpole diagrams both on the external quark and antiquark
propagators (i.e. self-energy contributions), and those coming from the
bilinear. In the latter case we include all diagrams in which the two
gluons both come from the bilinear, irrespective of whether they
emanate from the same smeared link. Thus, for the example of a distance-2
bilinear, which involves an average of a sum of products of two links,
the gluon can couple between the links (as well as going from each
link back to itself).

It is convenient to divide the contribution into two parts,
\begin{equation}
T_i = T_i^a + T_i^b
\,,
\end{equation}
with the former coming from gluon loops beginning and ending
on the same smeared link, and the latter involving gluons propagating
between smeared links.
In both cases the result depends only on the distance $\Delta$ of the bilinear.

No simple general formula covers all choices of links and action,
so we quote the results in turn.
\begin{enumerate}
\item
For unimproved staggered fermions, the result is (PS)
\begin{equation}
T^a_\Delta = (\Delta-1) \left(\pi^2 - \int_k B/2\right)
\,,
\label{eq:stagtad}
\end{equation}
where the factor of $\pi^2$ comes from tadpole improvement
using the fourth-root of the plaquette.
If one uses the trace of the Landau gauge link, then
$\pi^2$ is replaced by $\int_k 3 B/8$.
\item
For the Fat-7 links (without the Naik term), the result is
the same as for unimproved staggered fermions,
eq.~(\ref{eq:stagtad}), due to cancellations.
\item
For $O(a^2)$ improved links (but without the Naik term),
we find
\begin{equation}
T^a_\Delta = (\Delta-1) \left[
\left(\pi^2 - \int_k B/2\right)
+{3\over 2} \int_k B c_1 \bar s_2^2 \right]
\,.
\end{equation}
\item
For HYP links, we find
\begin{equation}
T^a_\Delta = (\Delta-1) \int_k (-B/2) P_{1}
\,,
\label{eq:HYPtad}
\end{equation}
where $P_\mu$ contains no Naik vertices:
\begin{equation}
P_{\mu}(k) = D_\mu(k)^2 + \sum_{\nu\ne\mu} G_{\nu,\mu}(k)^2
\,.
\end{equation}
We emphasize that, at this stage,
there is no tadpole improvement factor for the HYP links 
(although a related mean-field improvement will be introduced
in sec.~\ref{sec:tadhyp}).
It is also noteworthy
that this result would apply for both the Fat-7 and $O(a^2)$ improved links
were one to also include projection back into SU(3) in those 
cases~\cite{LSinprep}.
\item
Finally, for the Asqtad-like action we find
\begin{equation}
T^a_\Delta = (\Delta-1) \left[
\left(\pi^2 - \int_k B/2\right)
+{3\over 2} \int_k B c_1 \bar s_2^2 \right]
+ {1\over 4}\left[\pi^2 - \int_k B c_1 (1+c_1)\right]
\,,
\end{equation}
in which the second contribution is due to the Naik term.
\end{enumerate}

Now we turn to the ``off-diagonal'' tadpoles. These only arise from the bilinears,
and not from the self-energy contributions, and are only present for operators
with $\Delta\ge2$. Since they are off-diagonal they 
are not affected by $SU(3)$ projection, and so
take a common form for all actions and operators:
\begin{eqnarray}
T_{\Delta}^b &=& 4\int_k B \bar s_1^2 \bar s_2^2 O_{12} V_T(\Delta)
\\
V_T(2) &=& 1 \,,\ \
V_T(3) = 2 + c_3 \,,\ \
V_T(3) = 3 + 2 c_3  + c_3 c_4\,.
\end{eqnarray}
Here $O_{12}$ does not contain Naik contributions, even for the
Asqtad-like action:
\begin{equation}
4 \bar s_\mu \bar s_\rho O_{\mu\rho}(k) \ \equiv\ 
D_\mu G_{\mu,\rho} + D_\rho G_{\rho,\mu}
+ \sum_{\nu\ne(\mu,\rho)} G_{\nu,\mu}G_{\nu,\rho}
\,.
\end{equation}

\subsection{Self-energy diagrams}
\label{subsec:Z}

If the Naik term is not included in the action, the flavor-singlet
vector current is conserved, and it follows from the corresponding
Ward Identity that its matching factor vanishes. Thus it must be
that, in cases 1-4,
\begin{equation}
Z = - X_{ii} - Y_1\,,\qquad i=\opergi{\mu} \,.
\label{eq:zres}
\end{equation}
Here we have used the result that $T_1=0$ for cases 1-4.
We have checked eq.~(\ref{eq:zres}) analytically and numerically.

We cannot use this relation for the Asqtad-like action, 
since the hypercube vector operator is not the
conserved current (as it does not contain a Naik-like contribution).
A direct calculation is needed, and we find:
\begin{eqnarray}
Z &=& \int_k B^2 + \int_k B F I_Z \,,\\
I_Z &=& c_1 (1 + d_N s_1^2/2)(1 - 2 (s_1^N)^2 F)
\left(\bar c_1^2 P_1^{NN} - 3 \bar c_2^2 P_2^{NN}\right) \nonumber \\
&&\mbox{}
- s_1 s_1^N \left( 
P_1 + d_N (2 s_1^2/3 - \bar c_1^2) D_1 
- d_N s_1^2 \bar c_1^2 c_1/6 \right)\nonumber \\
&&\mbox{}
- 12 s_1 s_1^N c_1 (1 + d_N s_1^2/2) s_2 s_2^N F O_{12}^{NN} \nonumber \\
&&\mbox{}
- 3 \bar s_1^2 s_2 s_2^N \left(
4 O_{12} - d_N c_1 \bar c_1^2 \widetilde G_{1,2} 
+ d_N s_2^2 \widetilde G_{2,1}/6 \right) \,.
\end{eqnarray}
The first term in $Z$ is the standard integral used to subtract the
divergent piece. We have inserted $d_N$ in appropriate places
so that this result is valid for all the actions we consider.

\section{Further mean-field improvement of operators}
\label{sec:tadhyp}

It is possible to apply another level of tadpole, or, more accurately,
mean-field, improvement to the operators and actions which
involve smeared links.
Actually, for the HYP smeared links, this is the first level of tadpole 
improvement.
The fluctuations in the smeared links are reduced compared to those of the
original links but are still present. The residual fluctuations
can be estimated and partially removed by defining a smeared mean-link 
by analogy with the definition of the original mean-link~\cite{LM}:
\begin{equation}
(u_0^{SM})^{4} = \langle \mathrm{Smeared-Plaquette} \rangle
\,.
\end{equation}
Here the ``smeared-plaquette'' means the plaquette built out of smeared
links. The operators are then mean-field improved by
multiplying them by 
\begin{equation}
(u_0^{SM})^{1-\Delta}
\,,
\end{equation}
where $\Delta$ is the number of links in the bilinear.
The argument leading to this factor is identical to that used in PS
when tadpole improving staggered operators, and we do not repeat it here.
This procedure should be simple to implement in practice.

We have calculated the effect of such a mean-field improvement
for Fat-7, $O(a^2)$ improved, and HYP links. We have not applied it
to the case of the Asqtad-like action, because it is not entirely clear to
us how to incorporate the Naik term. 
We find that one must add to the tadpole contribution the following:
\begin{equation}
T_\Delta^c = (\Delta - 1) \left[ -T^a_{\Delta=2} - T^b_{\Delta=2}
-\int_k B P_1 c_2/2 \right]
\,,
\label{eq:Tc}
\end{equation}
where $T^{a,b}_\Delta$ and $P_1$ are given in the previous section.
The quantity in square parentheses evaluates to 
$-0.9125$ for Fat-7 links, 
$-4.0634$ for $O(a^2)$ improved links,
$0.5782$ for  HYP links with $\alpha_{1-3}=0.75,0.6,0.3$, and to
$1.0538$ for HYP links with the ``Fat-7 choice''
$\alpha_{1-3}=7/8,4/7,1/4$.
These values are substantially smaller than the analogous factor
in the tadpole improvement of the unsmeared links, namely $\pi^2$. 
They are, nevertheless, significant, as we see in the next section.

It is noteworthy that the Fat-7 and $O(a^2)$ improved smeared links receive
a mean-field correction of opposite sign to that of both the HYP links.
 This indicates that the fluctuations in the former case
have been ``overcompensated'' by smearing, and suggests that this higher
level of mean-field improvement is likely to be more significant for the
HYP smeared links.

Finally, we note that after this higher level of mean-field improvement,
the results for Fat-7 and HYP links with $\alpha_{1-3}=7/8,4/7,1/4$ are
identical. The equality of the tadpole contributions
can be seen by combining eqs.~(\ref{eq:stagtad}), (\ref{eq:HYPtad}) 
and (\ref{eq:Tc});
that of other contributions follows from the fact
that the single-gluon vertex is the same in both cases.

\section{Numerical results and discussion}
\label{sec:numerical}

We present numerical results for the matching coefficients 
in Tables~\ref{tab:cii}--\ref{tab:ciitad}.
As explained in Ref.~\cite{SP}, the corrections are unchanged if the
operators are multiplied by $\opergx55$, due to the conserved axial symmetry.
Thus we show results for only half the operators.
Recall that we have chosen the NDR scheme 
($\overline{MS}$ with an anticommuting
$\gamma_5$) and set $\mu=1/a$. We expect this to be a reasonable choice for the
matching scale, but, in any case, the dependence of the $c_{ij}$ on $\mu$ is weak,
as can be seen from eq.~(\ref{eq:matching}).\footnote{%
Approximate methods of calculating the optimal matching scale, $q^*$,
do not obviously generalize to the 
case of operators with non-vanishing anomalous
dimensions. We return to this issue elsewhere~\cite{LSinprep}.}

The most striking result from the tables is the significant reduction in
the size of one-loop corrections for all of the choices of smeared links.
This is true also for the off-diagonal matching constants, although here the
corrections were small to start with.
We also see that the mean-field improvement of Section~\ref{sec:tadhyp}
leads to a significant further reduction in the corrections for HYP
smearing, although the corrections increase somewhat
for Fat-7 and $O(a^2)$ improved smearing.

To compare the different alternatives for improvement we quote, in
Table~\ref{tab:ciirange}, the range of variation of the diagonal coefficients 
$c_{ii}$, both
for a given spin (varying the flavor), and for all spins and flavors. 
The range for a given spin is independent of
the renormalization scale $\mu$ (since $d_i$ is the same for all flavors), 
and thus is a good measure of the size of lattice contributions to matching
factors.
The range for all spins and flavors does depend on $\mu$, but only rather weakly.
The table shows that, of the alternatives we have compared, Fat-7 links,
with or without mean-field improvement,
and mean-field improved HYP or Fat-7 links lead to the smallest range of corrections.
A similar conclusion holds if we consider the maximum magnitude of the
corrections rather than the spread.

\begin{table}
\begin{center}
\begin{tabular}{crrrrrr}
Operator		      	&
	$(a)$ \quad &$(b)$ \quad &$(c)$ \quad &$(d)$ \quad &$(e)$\quad &$(f)$\quad \\ 
\hline 
$\operii$			&-29.3551& 1.8696&-4.3917&-2.1750&-0.5939&-0.0966\\
$\operix \mu$			& -8.6416& 2.4633&-2.5643&-0.3301& 1.8394& 2.4633\\
$\operix{\mu\nu}$		&  0.5657& 2.8990&-2.8420&-0.7999& 4.0139& 4.8653\\
$\operix{\mu5}$			&  5.2378& 3.3351&-4.0469&-2.1427& 6.0380& 7.2676\\
$\operix{5}$			&  8.7493& 3.7704&-5.5793&-3.7774& 7.9837& 9.6693\\
$\opergi{\mu}$			&  0.0000& 0.0000& 0.0000& 1.4155& 0.0000& 0.0000\\
$\opergx\mu\mu$			& -4.9092& 0.7869& 2.9240& 4.2755&-0.9457&-1.1794\\
$\opergx\mu\nu$			&  0.1721&-0.1201&-2.9799&-1.5110& 1.3090& 1.8461\\
$\opergx\mu{\mu\nu}$		& -3.3948& 0.3636&-0.0621& 1.4290& 0.2617& 0.3636\\
$\opergx\mu{\nu\rho}$		&  2.5040&-0.1930&-5.4907&-4.0295& 2.7140& 3.7396\\
$\opergx\mu{\nu5}$		&  0.1902& 0.1367&-2.5010&-1.0264& 1.6009& 2.1030\\
$\opergx\mu{\mu5}$		&  4.8930&-0.2147&-7.9437&-6.4957& 4.1592& 5.6841\\
$\opergx\mu{5}$			&  2.7709& 0.0369&-5.0332&-3.5799& 2.9898& 3.9694\\
$\opergi{\mu\nu}$		&  1.5969& 0.3741&-1.3115&-0.0393& 1.9442& 2.3404\\
$\opergx{\mu\nu}\mu$		&  0.8194& 0.8758& 2.1260& 3.3442& 0.9819& 0.8758\\
$\opergx{\mu\nu}\rho$		&  3.0150& 0.0410&-4.4862&-3.1752& 3.0313& 3.9735\\
$\opergx{\mu\nu}{\mu\nu}$	&  4.5728& 1.7594& 6.6960& 7.7590& 0.2703&-0.2069\\
$\opergx{\mu\nu}{\mu\rho}$	&  1.2809& 0.3800&-1.4041&-0.1178& 1.9309& 2.3463\\
$\opergx{\mu\nu}{\rho\sigma}$	&  4.9409&-0.2098&-7.3985&-6.0684& 4.2177& 5.6890\\
\end{tabular}
\end{center}
\caption[diagcoefs]{\fc
Diagonal part of the one-loop matching constants,
$c_{ii}$, using the NDR scheme with $\mu_{\rm NDR}=1/a$ in the continuum.
The components $\mu$, $\nu$, $\rho$ and $\sigma$ are all different.
Results are given for six choices of action and operators:
$(a)$ unimproved;
$(b)$ Fat-7 links;
$(c)$ fully $O(a^2)$ improved links;
$(d)$ fully $O(a^2)$ improved links and Naik term (Asqtad-like action);
$(e)$ HYP links with the smearing coefficients from Ref.~\cite{HK},
$\alpha_{1-3}=0.75,0.6,0.3$;
$(f)$ HYP links with tree-level improvement coefficients,
$\alpha_{1-3}=7/8,4/7,1/4$.
The error in the results is no larger than $0.0001$.}
\label{tab:cii}
\end{table}

\begin{table}
\begin{center}
\begin{tabular}{cccrrrrr}
Name & Operator-$i$   	& Operator-$j$ &
	$(a)$ \quad &$(b)$ \quad &$(c)$ \quad &$(d)$ \quad &$(e)$\quad \\ 
\hline 
$c_{VVM}$ &	$\opergx\mu\nu$			& $\opergx\mu\mu$	& 
 3.0412& 0.3508& 1.4104& 1.2976& 0.4203 \\
$c_{VAM}$ &	$\opergx\mu{\mu5}$		& $\opergx\mu{\nu5}$	& 
-0.6463&-0.2565&-0.6192&-0.5481&-0.3000 \\
$c_{VTM}$ &	$\opergx\mu{\mu\nu5}$		& $\opergx\mu{\rho\nu5}$& 
-1.4861&-0.2797&-0.9241&-0.8211&-0.3512 \\
$c_{TAM}$ &	$\opergx{\mu\nu}{\mu5}$		& $\opergx{\mu\nu}{\rho5}$& 
-0.6763& 0.0065&-0.2055&-0.1753&-0.0202 \\
\end{tabular}
\end{center}
\caption[diagcoefs]{\fc
Non vanishing off-diagonal one-loop matching constants, $c_{ij}$.
The components $\mu$, $\nu$ and $\rho$ are all different, but otherwise
can take any values.
Results are given for same choices of action and operators as in 
Table~(\protect\ref{tab:cii}), except that HYP links with tree-level coefficients
give identical results to Fat-7 links and thus are not shown.
The error in the results is no larger than $0.0001$.}
\label{tab:cij}
\end{table}

\begin{table}
\begin{center}
\begin{tabular}{crrr}
Operator		      	
	&$(b',f')$&$(c')$ &$(e')$\quad \\ 
\hline 
$\operii$			& 0.9571&-8.4551&-0.0156 \\
$\operix \mu$			& 2.4633&-2.5643& 1.8394 \\
$\operix{\mu\nu}$		& 3.8115& 1.2214& 3.4357 \\
$\operix{\mu5}$			& 5.1600& 4.0799& 4.8815 \\
$\operix{5}$			& 6.5079& 6.6110& 6.2490 \\
$\opergi{\mu}$			& 0.0000& 0.0000& 0.0000 \\
$\opergx\mu\mu$			&-0.1255&-1.1394&-0.3675 \\
$\opergx\mu\nu$			& 0.7924& 1.0835& 0.7308 \\
$\opergx\mu{\mu\nu}$		& 0.3636&-0.0621& 0.2617 \\
$\opergx\mu{\nu\rho}$		& 1.6320& 2.6361& 1.5576 \\
$\opergx\mu{\nu5}$		& 1.0492& 1.5624& 1.0227 \\
$\opergx\mu{\mu5}$		& 2.5227& 4.2466& 2.4245 \\
$\opergx\mu{5}$			& 1.8619& 3.0936& 1.8334 \\
$\opergi{\mu\nu}$		& 1.2866& 2.7519& 1.3660 \\
$\opergx{\mu\nu}\mu$		& 0.8758& 2.1260& 0.9819 \\
$\opergx{\mu\nu}\rho$		& 1.8659& 3.6407& 1.8748 \\
$\opergx{\mu\nu}{\mu\nu}$	& 0.8469& 2.6325& 0.8486 \\
$\opergx{\mu\nu}{\mu\rho}$	& 1.2925& 2.6593& 1.3527 \\
$\opergx{\mu\nu}{\rho\sigma}$	& 2.5276& 4.7918& 2.4830 \\
\end{tabular}
\end{center}
\caption[diagcoefs]{\fc
Results for $c_{ii}$ after the mean-field improvement
discussed in sec.~\protect\ref{sec:tadhyp}.
Notation as in Table~(\protect\ref{tab:cii}).
Results are given for 
$(b',f')$ Fat-7 links and
HYP links with tree-level improvement coefficients,
$\alpha_{1-3}=7/8,4/7,1/4$;
$(c')$ fully $O(a^2)$ improved links;
$(e')$ HYP links with smearing coefficients from Ref.~\cite{HK},
$\alpha_{1-3}=0.75,0.6,0.3$.}
\label{tab:ciitad}
\end{table}

\begin{table}
\begin{center}
\begin{tabular}{crrrrrrrrr}
Spin		      	&
	$(a)$  &$(b)$ &$(c)$ &$(d)$ &$(e)$ &$(f)$ & $(b',f')$ & $(c')$ & $(e')$  \\ 
\hline 
$S/P$	& 38.1& 1.9& 3.0& 3.4	& 8.6 	& 9.8	&5.5	&15.1	& 6.3	\\
$V/A$	&  9.8& 1.0&10.9&10.8	& 5.1	& 6.9 	&2.6	&5.4	& 2.8	\\
$T$	&  4.1& 2.0&14.1&13.8	& 3.9	& 5.9	&1.7	&2.7	& 1.6	\\
All	& 38.1& 4.0&14.6&14.3	& 8.9	&10.8	&6.6	&15.1	& 6.6	\\
\end{tabular}
\end{center}
\caption[diagcoefs]{\fc
Spread of values for diagonal corrections, $c_{ii}$, both for a given spin,
and between all operators. Notation as in Tables~(\protect\ref{tab:cii})
and (\protect\ref{tab:ciitad}).
}\label{tab:ciirange}
\end{table}

What values of $c_{ii}$ give rise to ``small enough'' corrections in present
simulations? Taking $1/a=2\,$GeV as a typical lattice spacing,
and using $\alpha_{\overline{MS}}(2{\rm GeV})\approx 0.19$, we find
$C_F \alpha_{\overline{MS}}/(4\pi)\approx 0.02$. 
Thus a matching coefficient $c=5$ corresponds to about a 10\% correction
at this lattice spacing. This is the size of corrections we are aiming for,
and we see that tadpole improved HYP fermions lead to corrections of
about this size.

Finally, it is interesting to compare to the size of one-loop corrections
for bilinears obtained with other fermion actions.
For unimproved Wilson fermions one finds, after tadpole improvement
(picking for definiteness the tadpole improvement scheme of Ref.~\cite{BGS}),
$c_i=-0.1,-9.7, -7.8, -2.9, -4.3$ for $i=S,P,V,A,T$, using the same
renormalization scheme and scale for the continuum operator as in the tables.
We have not been able to find the corresponding results for improved
Wilson fermions incorporating tadpole improvement,
but it is clear from 
Table 3 of Ref.~\cite{Capitani} that one-loop matching factors are of similar
size as for unimproved Wilson fermions.
For domain-wall fermions, tadpole-improved results 
are given in Ref.~\cite{Aoki}:
$c_i=-11.2, -5.3, -2.0$ for $i=S/P, V/A, T$ 
(setting the domain-wall mass $M=1.7$).
We conclude that the size of corrections with improved staggered fermions
is comparable to, or smaller than, that for other fermions.
This provides further impetus to pursue calculations with
improved staggered fermions.

\section*{Acknowledgments}
We thank Chris Dawson, Anna Hasenfratz and Francesco Knechtli for useful conversations.
The work of SRS is supported in part by the Department of Energy
through grant DE-FG03-96ER40956/A006.
The work of WL is supported in part by the BK21 program at Seoul National
University and in part by Korea Research Foundation (KRF) through
grant KRF-2002-003-C00033.

\newpage

\appendix
\section{Notation and technical details}
\label{sec:app}

We use the hypercube fields and bilinears introduced in Ref.~\cite{Kluberg}.
The lattice is divided into $2^4$ hypercubes
labeled by a vector $y$, with all components even.
Points within a hypercube are labeled by a hypercube vector
($C_\mu$, $D_\mu$ in the following), with all components $0$ or $1$.
The lattice bilinears we use are specified by ``spin'' and ``flavor''
hypercube vectors $S_\mu$ and $F_\mu$ in the following way in terms of
the staggered fields $\chi$ and $\bar\chi$:
\begin{equation}
{\cal O}_{\opergx{S}{F}}(y) = \frac1{16}\sum_{C,D}
\overline \chi^1(y+C) 
\sfno{S}{F}_{CD}
\chi^2(y+D)\,.
\label{eq:bilindef}
\end{equation}
Here the sums run over all positions in the hypercube,
and  
\begin{equation}
\sfno{S}{F}_{CD} =\frac14 \Tr\left[
\gamma_C^\dagger \gamma_S \gamma_D \gamma_F^\dagger\right]
\,,
\end{equation}
with
\begin{equation}
\gamma_S=\gamma_1^{S_1}\gamma_2^{S_2}\gamma_3^{S_3}\gamma_4^{S_4}
\end{equation}
composed of hermitian Euclidean gamma matrices.
It follows that the $\bar\chi$ and $\chi$ fields are separated by
a fixed number of links which is given by the ``distance''
$\Delta = \sum_\mu (S_\mu - F_\mu)^2$.
In the continuum limit, this lattice bilinear has the same spin, flavor 
and normalization as the continuum bilinear
\begin{equation}
{\cal O}^{\rm cont}_{\opergx{S}{F}} =
\overline Q^1_{\alpha, a} \gamma_S^{\alpha\beta} \xi_F^{ab} Q^2_{\beta, b}
\,,
\end{equation}
where
$Q^k_{\beta, b}$ is a four-flavor quark field, with
spinor index $\beta$ and flavor index $b$, both running from $1$ to $4$,
and the
$\xi_F=\gamma_F^*$
form a convenient basis for the flavor matrices.

The superscript on $\chi^k$, $Q^k$, etc. indicates
an additional flavor index, corresponding to the different continuum
flavors ($u$, $d$, $s$, etc.). We consider here only continuum flavor non-singlet
operators, so that we do not have to calculate diagrams in which the fermion
fields in the bilinear are contracted together.

Because the lattice bilinears are spread over a hypercube, the
phase factors due the external quark and antiquark momenta
depend upon the hypercube vectors $C$ and $D$ in eq.~(\ref{eq:bilindef}).
To disentangle these phases Daniel and Sheard introduced 
the following definition~\cite{DS}:
\begin{equation}
e^{ik\cdot C} =
\left(\prod_\mu e^{i k_\mu/2}\right)
\sum_M E_M(k) (-)^{C\cdot \widetilde M}
\end{equation}
with
\begin{equation}
E_M(k) = \prod_\mu \frac12
\left(e^{- i k_\mu/2} + (-)^{\widetilde M_\mu} e^{i k_\mu/2} \right)
\,.
\end{equation}
Here we have introduced the conjugate hypercube vectors
\begin{equation}
\widetilde M_\mu =_2 \sum_{\nu\ne\mu} M_\nu
\,.
\end{equation}
Previous results for $X$ diagrams in Ref.~\cite{DS} and PS were expressed as
sums over the additional hypercube vector $M$, which were then evaluated
numerically.
We have been able to perform this sum analytically, which simplifies
the final expressions.
The two results we need in this paper are:
\begin{equation}
V_i(k)
\equiv
\sum_M E_M(k) E_M(-k) (-)^{M\cdot(\widetilde S+ \widetilde F)}
= \prod_\mu \cos[k_\mu(S\!-\!F)_\mu]
\,,
\label{eq:vertex}
\end{equation}
where $i$ labels the spin and flavor,  and
\begin{equation}
\sum_M E_M(k) E_{M+\mu+\nu}(-k) (-)^{M\cdot(\widetilde S+ \widetilde F)}
= -|(S\!-\!F)_\mu||(S\!-\!F)_\nu| s_\mu s_\nu
\cos[k_\rho(S\!-\!F)_\rho] \cos[k_\sigma(S\!-\!F)_\sigma]
\,,
\label{eq:mixingvertex}
\end{equation}
in which all indices are different.
These results can be obtained by combining PS eqs. (24) and (27).

The explicit forms for lattice integrals are abbreviated using the
following notations:
\begin{equation}
\int_k \equiv 16\pi^2 \prod_\mu \int_{-\pi}^{\pi} {dk_\mu \over 2\pi}\,,\quad
\bar s_\mu \equiv \sin(k_\mu/2)\,,\ 
s_\mu \equiv \sin(k_\mu)\,,\
\bar c_\mu \equiv \cos(k_\mu/2)\,,\ 
c_\mu \equiv \cos(k_\mu)\,.
\end{equation}

\end{document}